\documentstyle[twoside,fleqn,espcrc2,epsfig]{article}

\def\beq{\begin{equation}}
\def\eeq{\end{equation}}
\def\beqa{\begin{eqnarray}}
\def\eeqa{\end{eqnarray}}
\def\be{\begin{equation}}
\def\ee{\end{equation}}
\def\bea{\begin{eqnarray}}
\def\eea{\end{eqnarray}}
\newcommand{\AmS}{{\protect\the\textfont2
   A\kern-.1667em\lower.5ex\hbox{M}\kern-.125emS}}

\title{
$K \rightarrow \pi \pi$  Electroweak Penguins in 
the Chiral Limit
}

\author{V. Cirigliano \address{Departament de F\'isica Te\`orica, 
           IFIC, Universitat de Val\`encia - CSIC , \\
          Apt. Correus 22085, E-46071 Val\`encia, Spain}%
         \thanks{Speaker at QCD02. 
          Research  supported in part by EC-Contract ERBFMRX-CT980169, 
          by MCYT, Spain (Grant No. FPA-2001-3031), and by ERDF funds from 
          the European Commission} ,        
         J.F. Donoghue and E. Golowich   
        \address{Physics Department, University of Massachusetts, \\
            Amherst MA 01003  USA } 
         \thanks{Research supported in part by the NSF under Grant 
          PHY-9801875}, 
         K. Maltman 
         \address{Dept. of Mathematics and Statistics, 
             York University, 
            4700 Keele St., Toronto ON M3J 1P3 Canada,  \\
	   and CSSM, University of Adelaide, Adelaide 5005, SA, Australia}
  \thanks{Research supported in part by the Natural Science and 
           Engineering Research Council of Canada }
}

\begin{document}

\begin{abstract}

We report on  dispersive and finite energy sum rule analyses 
of the electroweak penguin
matrix elements $\langle (\pi \pi)_2 | {\cal Q}_{7,8} | K^0 \rangle $
in the chiral limit.  We accomplish the correct perturbative
matching (scale and scheme dependence) at NLO in
$\alpha_s$, and we describe two different strategies for  
numerical evaluation.  
\end{abstract}
\maketitle

\section{Introduction}
\label{sect:intro}

The subject of this talk is  the calculation of the  
$K \rightarrow \pi \pi$  matrix elements of the electroweak penguin 
operators, conventionally denoted as ${\cal Q}_{7}$ and 
${\cal Q}_{8}$ ~\cite{buras1}. 
%\beqa 
% {\cal Q}_{7} & = & {3 \over 2} \ \bar{s} ( \Gamma_L  )_{\mu} 
%d \, \times \,  \bar{q} \,   Q   \, (\Gamma_{R}  )^{\mu}  
%  q   
%%\nonumber  
%\\ 
%{\cal Q}_{8} & = & {3 \over 2} \   \bar{s}_{\alpha} ( \Gamma_L  )_{\mu}
% d_{\beta} \, \times \,  \bar{q}_{\beta} \,  Q  \,  
%( \Gamma_{R}  )^{\mu}  q_{\alpha}  
%%\nonumber 
%\eeqa
%where $q$ indicates the flavor-vector of active quark fields,   
%$Q$ is their electric charge matrix, and 
%$\Gamma_{L,R} =  1 \mp \gamma_5$. 
Interest in these quantities is twofold. First, the Standard Model
prediction of $\epsilon'/\epsilon$ depends crucially on the
K-to-$2\pi$ matrix elements of the QCD penguin (${\cal Q}_6$) and EW
penguin (${\cal Q}_8$) operators,~\cite{buras1,Bertolini,ciuchini}:
$$ {\epsilon' \over \epsilon} = 20 \cdot 10^{-4}
%\left( { \IM V_{ts}^* V_{td}  \over 1.3 \cdot 10^{-3}}  \right)
\bigg[ 
%- 0.06   
%$$ 
%\vspace*{0.2cm}
%$$ 
- 2.0 \cdot \frac{  \langle {\cal Q}_6\rangle_0
%^{\scriptstyle (2~{\rm GeV})}
 }{{\rm GeV}^{3}} \, (1 - \Omega_{\rm IB}) 
- 0.50 \cdot \frac{  \langle {\cal Q}_8\rangle_2
%^{\scriptstyle (2~{\rm GeV})} 
 }{{\rm GeV}^{3}}   \bigg]  , $$
where the above matrix elements are evaluated in the ${\overline
{MS}}$-NDR renormalization scheme at scale $\mu =
2$~GeV.  Secondly, the leading term in the chiral expansion of 
$ \langle {\cal Q}_{7,8} \rangle $ 
can be related to the $V - A$ QCD correlator with flavor $ud$, whose
spectral function is experimentally accessible through $\tau$ decays 
\cite{aleph,opal}. This provides the 
opportunity to perform a data-driven evaluation of hadronic matrix
elements, whose results can then be confronted with other
non-perturbative techniques (like Lattice QCD and $1/N_c$ expansion).
Therefore electroweak penguins provide an interesting theoretical
laboratory for kaon physics.

\section{Summary of formal results}

Our first goal is to relate the $K \rightarrow \pi \pi$ matrix elements 
to the observable spectral functions. The first large step is done for us 
by chiral symmetry, which relates  $ \langle \pi \pi | {\cal Q}_{7,8}
 | K \rangle $ to vacuum expectation values of two  4-quark 
operators~\cite{dg99}
\beqa
\langle (\pi\pi)_{2} | {\cal Q}_7
|K^0\rangle_{p^0} & = & - {2 \over F_{0}^3 }~\langle {\cal O}_1
\rangle \ \ , 
\nonumber 
\\
\langle (\pi\pi)_{2} | {\cal Q}_8 | K^0\rangle_{p^0} & = & -
{2 \over F_{0}^3} ~\left[ {1 \over 3} \langle {\cal O}_1 \rangle + {1
\over 2} 
\langle {\cal O}_8 \rangle \right]  , \qquad \qquad
\nonumber 
\eeqa 
where $F_{0}$ is the pion decay constant in the chiral limit and 
the definitions of $O_1$ and $O_8$ can be found in Ref.~\cite{dg99}.
%\footnote{
%The operators
%${\cal O}_{1,8}$ are defined as
%\begin{eqnarray}
%{\cal O}_1 &\equiv& {\bar q} \gamma_\mu {\tau_3 \over 2} q
%~{\bar q} \gamma^\mu {\tau_3 \over 2} q -
%{\bar q} \gamma_\mu \gamma_5 {\tau_3 \over 2} q
%~{\bar q} \gamma^\mu \gamma_5 {\tau_3 \over 2} q \ \ ,
%\nonumber \\
%{\cal O}_8 &\equiv& {\bar q} \gamma_\mu \lambda^a
%{\tau_3 \over 2} q
%~{\bar q} \gamma^\mu \lambda^a {\tau_3 \over 2} q -
%{\bar q} \gamma_\mu \gamma_5 \lambda^a {\tau_3 \over 2} q
%~{\bar q} \gamma^\mu \gamma_5 \lambda^a {\tau_3 \over 2} q .
%\nonumber 
%\end{eqnarray}
%}
The above relations receive NLO chiral corrections
(whose order of magnitude is $\sim m_K^2/(4 \pi F_\pi)^2 = 0.18$)
which have been investigated within chiral 
perturbation theory~\cite{cg}. 

The second step consists in relating the v.e.v.'s of $O_{1,8}$ to the
$V-A$ QCD correlator $\Delta \Pi_{ud} (s) \equiv \Pi_{\rm V}^{(0+1)} (s) -
\Pi_{\rm A}^{(0+1)} (s)$, with flavor structure 
$ud$ \footnote{For details on the notation we refer to
Ref.~\cite{cdgm}}.  The outcome is that:
\begin{itemize}
\item  $\langle O_1 \rangle$ is 
related to the integral of $s^2 \Delta \Pi (s) $ over all spacelike 
momenta;
%while 
\item  $\langle O_8 \rangle$ is related to the leading
singularity of $\Delta \Pi (s) $ at short distance (large spacelike
momenta). 
\end{itemize}
Essential tools in a formal derivation of the above results 
and in relating $\langle O_{1,8} \rangle $ to the spectral function  
are the  dispersive representation of the correlator
\beq
\Delta \Pi (Q^2)= - {2 F_\pi^2 \over m_\pi^2 + Q^2} +
\int_{s_{\rm thr}}^\infty ds\ {\Delta \rho (s)  \over s + Q^2} 
\, .\label{r7}
\eeq
and the operator product expansion (OPE) representation, 
valid for large space-like momenta $Q^2 \gg \Lambda_{\rm QCD}^2$,  
which, through ${\cal O} (\alpha_s^2)$,  has the form  
\beq
\Delta \Pi_{OPE} (Q^2) \sim \displaystyle\sum_{d \geq 2} \ {1 \over Q^d} \left[
a_d (\mu) + b_d (\mu) \ln{Q^2 \over \mu^2} \right]  .
%\qquad (d = 2,4,6,8,10, \dots) \
\label{r11}
\eeq
$a_d (\mu)$ and $b_d (\mu)$ are combinations of vacuum
expectation values of local operators of dimension $d$.
In the chiral limit, the above sum begins with $a_6$, whose 
NLO expression is \footnote{The coefficients $A_{1,8}$ 
depend on the definition used for $\gamma_5$ in $d \neq 4$. 
They can be found in Ref.~\cite{cdgm}.}:
\beq
a_6 (\mu) = (2 \pi \alpha_s + A_8 \alpha_s^2) \langle O_8 \rangle + 
    A_1 \alpha_s^2 \langle O_1 \rangle  \ . 
\eeq
The formal analysis is summarized by 
\beq
\left( \begin{array}{c} \langle O_1 \rangle_\mu \\ 
%  \\  
\langle O_8 \rangle_\mu \end{array} \right)  = 
\hat{M} (\alpha_s) \ 
\left( \begin{array}{c}  I_1 (\mu) + H_1 (\mu) \\  
%  \\ 
 I_8 (\mu) - H_8 (\mu)    
\end{array} \right) \ , 
\label{eq:basic1}
\eeq
where the matrix $\hat{M} (\alpha_s)$ contains 
the matching coefficients at NLO in $\alpha_s$, 
and encodes the correct renormalization scheme dependence. 
Its explicit form can be found in Refs.~\cite{cdgm}. 
$I_{1,8}$ are related to the spectral function as follows,  
\beqa
%\begin{array}{ccc}
  I_1 (\mu)  & = & 
\displaystyle  \int_0^\infty ds\ s^2 \ln
\left({s + \mu^2 \over s} \right) \Delta \rho (s)  
\\
%  & &  \\
  I_8 (\mu)   &  = &  
\displaystyle  \int_0^\infty ds\ {s^2 \mu^2 \over s + \mu^2} \ 
\Delta \rho (s)  
%\end{array}
\eeqa
while $H_{1,8}$ are related to subleading (higher dimensional) terms
in the short distance expansion:
% of $\Delta \Pi (Q^2)$:
\beqa
%\begin{array}{ccc}
  H_1 (\mu)  & = & 
\sum_{d \geq 8} \displaystyle 
{2 \over d-6} \cdot {a_d (\mu) \over \mu^{d-6}} \ , 
\\
%  & &  \\
  H_8 (\mu)   &  = &  
\sum_{d \geq 8} \displaystyle{a_d (\mu) \over \mu^{d-6}} \ .  
%\end{array}
\eeqa

A close look at the input needed in Eq.~(\ref{eq:basic1}) and at the
database on $\Delta \rho$ (Refs.~\cite{aleph,opal}) shows that
data alone do not provide enough information for a reliable
determination of the spectral integrals.  In fact, we run out of data
at $s=m_\tau^2$, while the weighs entering  $I_{1,8}$ are growing
with $s$.  We therefore need to supply some extra theoretical
input to optimize the use of data in this problem. Along this path, 
we have followed two approaches, which can be schematically 
characterized as follows: 
\begin{itemize}
\item[I.]
 Use of data plus QCD integral constraints (classical chiral sum 
 rules).~\cite{cdgm} 
\item[II.] Use of data plus QCD duality 
(Finite Energy Sum Rules, FESR).~\cite{cdgm2} 
\end{itemize}

\section{Numerics I: Residual Weight Method}

This approach is an efficient way to enforce the constraints from
classical chiral sum rules~\cite{weinberg,das} in the evaluation of
the spectral integrals $I_{1,8} (\mu)$, which have the generic form: 
\beq
I(\mu) = \int_0^\infty  ds \, K (s,\mu) \, \Delta \rho (s) \ . 
\eeq
The basic idea is to write $K(s,\mu)$ as the sum of 
an arbitrary combination of
the weights $1,s$, and $s \log s$ and a residual term $r$, as
follows:
\beq
K (s,\mu) = x \, 1 + y \, s + z \, s \log s  + r_{(x,y,z)} (s, \mu) \ . 
\label{rwm1}
\eeq
From Eq.~(\ref{rwm1}) and the chiral constraints one then has
\beq
I (\mu) = F_0^2 (x - z {4 \pi \delta M_\pi^2 \over 
3 \alpha } ) + \! \int_0^\infty  \! \! \! \! \! \! 
ds \,  r (s, \mu) \,  \Delta \rho (s) ,    
\eeq
and can choose $(x,y,z)$ (so far arbitrary) in such a way as to minimize
the total uncertainty (associated with both the theoretical input for 
 $F_0$ and $\delta M_\pi^2$,  and the large errors on 
 $\Delta \rho (s)$ for $s > 2.5$ GeV$^2$).  
The method is effective in those situations where 
choices for $x,y,z$ exist which make $r \sim 0$ in the region 
with no data (but before the onset of asymptotia) 
without amplifying the effect of imprecise knowledge 
of $F_0$ and $\delta M_\pi^2$. 
%Clearly the method has chances to work only if we can adjust
%$x,y,z$ to make $r \sim 0$ in the region with no data (before the
%asymptotic behavior sets in) without amplifying too much the effect of
%imprecise knowledge of $F_0$ and $\delta M_\pi^2$. 
Procedural details
and intermediate results can be found in Ref.~\cite{cdgm}.  Here let
us recall that the method leads to a reasonably accurate determination
$I_{1,8}$ at different values of $\mu$ between 2 and 4 GeV (with
accuracy deteriorating at large $\mu$).  This method tells us very
little about  the subleading terms $H_{1,8}$, which can be neglected only
for sufficiently large $\mu$. 
We have assumed that $H_{1,8}(\mu = 4 \mbox{GeV}) =0$
in Eq.~(\ref{eq:basic1}) 
and used the NLO renormalization group equations to evolve the results 
down to the lower scale  $\mu= 2$ GeV. 
Numerical results are reported in Sect.~\ref{sect:results}

\section{Numerics II: FESR}

In principle, this method allows us to determine the coefficients
$a_d$ appearing in the OPE expansion for the correlator $\Delta \Pi
(s)$ in the asymptotic Euclidean region  (Eq.~(\ref{r11})).  
The determination of $a_6$, in particular, provides a direct 
extraction of $\langle O_8 \rangle$
at $\mu=2$ GeV.  The determination of the $a_d$, $d \geq 8$, 
also allows one to estimate the combinations  $H_{1,8} (\mu = 2 \, 
 \mbox{GeV})$, needed as input to the lower scale version of 
the RWM analysis, Eq.~(\ref{eq:basic1}).  Let us now summarize our
implementation of the FESR analysis~\cite{cdgm2}. 

An FESR analysis, applied to a generic correlator $\Pi (s)$ with
spectral function $\rho (s)$ is based on the following relation 
(consequence of Cauchy's theorem, and the analytic structure of the 
correlator):
\beq
J[w, s_0] +  R[w, s_0] = f_{w} [s_0, \{ a_d \}] \ . 
\label{du3}
\eeq
Eq.~(\ref{du3}) involves the spectral integral 
\beq
J[w,s_0] = \int_{s_{th}}^{s_0} ds \ w(s)  \rho (s) \ , 
\eeq
the integrated OPE (function of the various
condensates) 
\beq
f_{w} [s_0, \{ a_d \}] = 
-\frac{1}{2 \pi i} \oint_{|s| = s_0} ds \  w(s) \Pi_{\mbox{\tiny OPE}} (s) \ , 
\eeq
and the remainder term (arising from the difference $\Pi (s) - \Pi_{OPE}
(s)$ on the $|s|=s_0$ circle) 
\beq
R[w,s_0] = 	
-\frac{1}{2 \pi i} \oint_{|s| = s_0} ds \  w(s) \left( 
 \Pi_{\mbox{\tiny OPE}} -  \Pi \right)  . 
\eeq
Eq.~(\ref{du3}) is valid for any value of $s_0$ and any 
polynomial weight. However, the presence of $R [s_0,w]$ pollutes the
extraction of the OPE coefficients $a_d$ in terms of data through 
Eq.~(\ref{du3}).  Therefore it is highly desirable to work with a
range of $s_0$ values and with weights $w(s)$ such that
 $R [w,s_0] << J[w,s_0]$.
To accomplish this, we rely on the observation that
for $|s|$ large enough ($s_0 \gg \Lambda_{QCD}^2$), the OPE
provides a good representation of the full correlator along the whole
circle, except in a region localized around the time-like axis.  The
physics of this breakdown is given by the arguments of Poggio, Quinn
and Weinberg \cite{poggio}.
As a consequence one expects that weights with a zero at $s=s_0$,
de-emphasizing the region where the OPE fails, are good candidates
to generate a small-sized $R [w,s_0]/J[w,s_0]$ \footnote{Supporting 
evidence for this fact has been found in Ref.~\cite{KM98}, within 
an analysis of the  V and A correlators.}.
We use this as the basic guiding principle in the 
selection of weights for our analysis. The weights were also 
chosen to avoid large cancellations in the spectral 
integrals, which would lead to a poor statistical signal. 

\begin{figure}[htb]
\begin{center}
\leavevmode
\begin{picture}(100,50)  
\put(10,5){\makebox(50,50){\epsfig{file=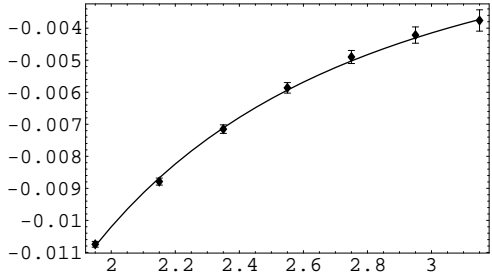,height=4.0cm}}}
\end{picture}
\caption{ $f_{w_1}[s_0,a_6,a_8]$ (continuous curve) and $J[w_1,s_0]$
(data points) versus $s_0$ (GeV$^2$).  }
\label{fig:w1}
\end{center}
\end{figure}

\begin{figure}[htb]
\begin{center}
\leavevmode
\begin{picture}(100,30)  
\put(10,5){\makebox(50,50){\epsfig{file=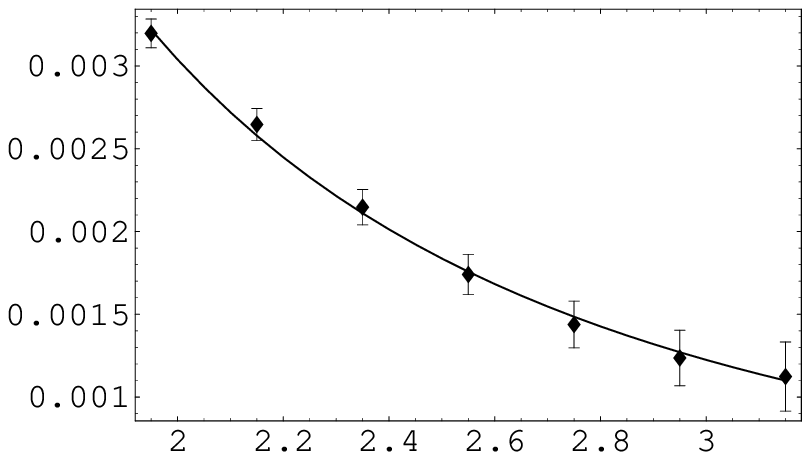,height=4.0cm}}}
\end{picture}
\caption{ $f_{w_2}[s_0,a_6,a_8]$ (continuous curve) and $J[w_2,s_0]$
 (data points) versus $s_0$ (GeV$^2$).  }
\label{fig:w2}
\end{center}
\end{figure}

The lowest degree weights satisfying our criteria 
have degree 3. Two useful cases (with $y=s/s_0$) are: 
\beqa
w_1 (y) &=& (1 - y)^2 \,  ( 1 - 3 y )  \\  
w_2 (y) &=& (1 - y)^2 \, y   
\eeqa
These choices imply that  $f_{w_{1,2}} [s_0, \{ a_d \}]$ depend on 
the dimension six and eight condensates, thereby allowing us to 
extract information on $a_6$ and $a_8$ from a least-square 
fit. The  window of $s_0$ values used in the fit procedure 
is $1.95 \, \mbox{GeV}^2 < s_0 <  3.15 \, \mbox{GeV}^2$. 
%The upper endpoint of the range is forced upon us, as the data sample
%for $\Delta\rho(s)$ ends there.  
The lower endpoint has been
determined by trying ever lower values of $s_0$ until 
the extracted $a_6$ ceases to be consistent with the 
values obtained by the smaller analysis windows. 
From the ALEPH data we obtain (at $\mu = 2$ GeV): 
\beqa
a_6  &=&
\left( - 4.45 \pm 0.7 \right) \cdot 10^{-3}~{\rm GeV}^6
\label{ope}  \\
a_8  &=& \left( - 6.2 \pm 3.2  \right)
\cdot 10^{-3}~{\rm GeV}^8 
\eeqa
with a correlation coefficient $c (a_6, a_8) = -0.99$.  
Although no precise answer emerges for $a_8$, we see 
that $a_6$ (directly related to ${\cal Q}_8$) can be 
determined fairly well. 
The errors reported above are essentially of statistical nature.  In
order to address potential systematic effects due to $R [s_0,w] \neq 0$
(duality violation), we have repeated the fit procedure 
using  non-overlapping $s_0$ sub-windows. We
find that the fitted parameters in the different analyses are consistent, 
thereby confirming that the effect of $R [s_0,w]$ is
suppressed in this case.  A more explicit portrait of
the FESR machinery is obtained by plotting $J [w_{1,2},s_0]$ and 
$f_{w_{1,2}} [s_0;a_6, a_8]$ as a function of $s_0$, 
as in Figs.~\ref{fig:w1},\ref{fig:w2}. 
The excellent match of the OPE curve versus data increases our 
confidence that in this analysis duality violation is under control. 
\vspace{-.5cm}
\vspace{-1.5cm}

\section{Results for $\langle (\pi \pi)_{I=2} | {\cal Q}_{7,8} | K^0 \rangle$}
\label{sect:results}  

Having outlined the two numerical strategies in the above sections, 
in Table \ref{tab:tab1}  we report $\langle {\cal Q}_{7,8} \rangle $ in the 
${\overline {MS}}$-NDR renormalization scheme at scale
$\mu =2 $ GeV, from methods I, II, and VSA 
\footnote{The number reported corresponds to 
$(m_s + m_d) (2 GeV) = 110$ MeV. }.
%
%\vspace{-0.5cm}
\begin{table}[htb]
\caption{${\overline {MS}}$-NDR results at $\mu = 2$ GeV and $n_f=4$.} 
\label{tab:tab1}
\begin{tabular}{ccc}
\hline 
Method  &  $\langle {\cal Q}_{7} \rangle_{I=2}$/GeV$^3$    & 
$\langle {\cal Q}_{8} \rangle_{I=2}$/GeV$^3$  \\
\hline 
 I &   0.16 $\pm$ 0.10 &  2.22 $\pm$ 0.70  \\
\hline 
% II & \  0.21 $\pm$ 0.03 &  1.62 $\pm$ 0.34 \\
 II (ALEPH)  & \  0.23 $\pm$ 0.05 &  1.41 $\pm$ 0.27 \\
 II (OPAL)  & \  0.21 $\pm$ 0.05 &  1.72 $\pm$ 0.32 \\
\hline  VSA  & 0.32  & 0.94 \\
\hline
\end{tabular}
\end{table}
%\vspace{-0.6cm}

Procedures I and II lead to consistent results, with II having reduced
uncertainty. The underlying reason is that method II exploits best
the low-error part of experimental data.  Both our results seem to
point to moderate violation of VSA.  For a comparison with other
recent determinations \cite{bgp01,kpr,nar,pps,latt} see
Refs.~\cite{bgp01,cdgm2}.  Implications of our results for the value
of $\epsilon ' / \epsilon$ in the Standard Model will be discussed
in~\cite{cdgm2}.


\begin{thebibliography}{9}


%\cite{Buras:1998ra}
\bibitem{buras1}
A.~J.~Buras, hep-ph/9806471.   

\bibitem{Bertolini}
S.~Bertolini,
%``Theory status of epsilon'/epsilon,'' 
hep-ph/0206095.
%%CITATION = HEP-PH 0206095;%%

%\cite{Ciuchini:1995qx}
\bibitem{ciuchini}
M.~Ciuchini et. al, 
%``Estimates of $\epsilon' / \epsilon$,'' 
hep-ph/9503277.
%%CITATION = HEP-PH 9503277;%%

\bibitem{aleph}
ALEPH Coll., R. Barate et al., Eur. Phys. J. C {\bf 4} 
(1998) 409.   

\bibitem{opal}
OPAL Coll., K. Ackerstaff et al.,  
Eur. Phys. J. C {\bf 7} (1999) 571. 


\bibitem{dg99}
J.~F.~Donoghue and E.~Golowich,
%%``Dispersive calculation of B7(3/2) and B8(3/2) in the chiral limit,''
Phys.\ Lett.\ B {\bf 478} (2000) 172.
%[arXiv:hep-ph/9911309].
%%CITATION = HEP-PH 9911309;%%

\bibitem{cg}
V.~Cirigliano and E.~Golowich,
%%``Analysis of O(p**2) corrections to ,''
Phys.\ Lett.\ B {\bf 475} (2000) 351, 
%[arXiv:hep-ph/9912513], 
%%CITATION = HEP-PH 9912513;%%
Phys.\ Rev.\ D {\bf 65} (2002) 054014. 
%[arXiv:hep-ph/0109265].
%%CITATION = HEP-PH 0109265;%%

\bibitem{cdgm}
V. Cirigliano et al.,  Phys. Lett. {\bf B 522} (2001) 245. 
%[hep-ph/0109113].      


\bibitem{cdgm2}
V. Cirigliano, J.F. Donoghue, E. Golowich and K. Maltman,  
 in preparation. 

\bibitem{weinberg}
S. Weinberg, Phys. Rev. Lett. {\bf 18} (1967) 507. 

\bibitem{das}
T. Das et al., Phys. Rev. Lett. {\bf 18} (1967) 759.


\bibitem{poggio}
E. Poggio et al., Phys. Rev. D {\bf  13} (1976) 1958.

\bibitem{KM98}
K. Maltman, Phys. Lett. {\bf B 440} (1998) 367. 

\bibitem{bgp01}
J.~Bijnens et al., 
JHEP {\bf 0110}, (2001) 009 ; and hep-ph/0209089.
%%CITATION = HEP-PH 0108240;%%

\bibitem{kpr}
M. Knecht et al., 
Phys.\ Lett.\ B {\bf 508} (2001) 117. 

\bibitem{nar}
S. Narison, Nucl. Phys. B {\bf 593} (2001) 3.

\bibitem{pps}
E. Pallante et al., Nucl.\ Phys.\ B {\bf 617} (2001) 441.
%%CITATION = HEP-PH 0105011;%%

\bibitem{latt}
A. Donini et al., 
Phys.\ Lett.\ B {\bf  470} (1999) 233 ; 
CP-PACS Coll., J.-I. Noaki et al., hep-lat/0108013 ;
RBC Coll., T. Blum et al., hep-lat/0110075 ;
SPQ$_{CD}$R Coll., D. Becirevic et al., hep-lat/0209136. 

\end{thebibliography}
\end{document}